\documentclass[final]{aipproc}
\layoutstyle{6x9}
\usepackage{amsmath,wrapft,epsfig}
\begin{document}

\begin{flushleft}
\scalebox{1.3}{SAGA-HE-198-03}   \hfill    
\scalebox{1.3}{October 1, 2003}  \\
\end{flushleft}
\vspace{2.6cm}

\begin{center}
\scalebox{1.7}{\bf Neutrino}   \hspace{0.0cm}
\scalebox{1.7}{\bf Scattering} \hspace{0.0cm}
\scalebox{1.7}{\bf Physics}    \\
\vspace{0.3cm}
\scalebox{1.7}{\bf at}         \hspace{0.0cm}
\scalebox{1.7}{\bf Superbeams} \hspace{0.0cm}
\scalebox{1.7}{\bf and}        \hspace{0.0cm}
\scalebox{1.7}{\bf Neutrino}   \hspace{0.0cm}
\scalebox{1.7}{\bf Factories}  \\
\vspace{1.5cm}
\scalebox{1.4}{\ S. Kumano $^*$} \\
\vspace{0.5cm}\scalebox{1.4}{Department of Physics}   \\
\vspace{0.2cm}\scalebox{1.4}{Saga University}         \\
\vspace{0.2cm}\scalebox{1.4}{Saga 840-8502, Japan}    \\

\vspace{1.8cm}
\scalebox{1.3}{Plenary}        \hspace{0.0cm}
\scalebox{1.3}{talk}           \hspace{0.0cm}
\scalebox{1.3}{at}             \hspace{0.0cm}
\scalebox{1.3}{the}            \hspace{0.0cm}
\scalebox{1.3}{5th}            \hspace{0.0cm}
\scalebox{1.3}{International}  \hspace{0.0cm}
\scalebox{1.3}{Workshop}       \hspace{0.0cm}
\scalebox{1.3}{on}             \\
\vspace{0.2cm}
\scalebox{1.3}{Neutrino}       \hspace{0.0cm}
\scalebox{1.3}{Factories}      \hspace{0.0cm}
\scalebox{1.3}{\&}             \hspace{0.0cm}
\scalebox{1.3}{Superbeams}     \\
\vspace{0.50cm}
\scalebox{1.3}{Columbia}       \hspace{0.0cm}
\scalebox{1.3}{University,}    \hspace{0.0cm}
\scalebox{1.3}{New York,}      \hspace{0.0cm}
\scalebox{1.3}{USA}            \\
\vspace{0.35cm}
\scalebox{1.3}{June}           \hspace{0.0cm}
\scalebox{1.3}{5-11,}          \hspace{0.0cm}
\scalebox{1.3}{2003}           \hspace{0.0cm}
\scalebox{1.2}{(talk}          \hspace{0.0cm}
\scalebox{1.2}{on}             \hspace{0.0cm}
\scalebox{1.2}{June}           \hspace{0.0cm}
\scalebox{1.2}{5)}             \\
\end{center}
\vspace{2.5cm}
\vfill
\noindent
{\rule{6.0cm}{0.1mm}} \\

\vspace{-0.3cm}
\normalsize
\noindent
\scalebox{1.1}
{* Email: kumanos@cc.saga-u.ac.jp. URL: http://hs.phys.saga-u.ac.jp.} \\

\vspace{+0.8cm}
\hfill
\scalebox{1.1}
{to be published in AIP proceedings}

\vfill\eject
\normalsize

\title{Neutrino Scattering Physics \\
       at Superbeams and Neutrino Factories}
\author{S. Kumano}
{address={Department of Physics, Saga University, Saga, 840-8502, Japan},
   email={kumanos@cc.saga-u.ac.jp}}

\begin{abstract}
Neutrino scattering physics is discussed for investigating internal
structure of the nucleon and nuclei at future neutrino facilities.
We explain structure functions in neutrino scattering. In particular,
there are new polarized functions $g_3$, $g_4$, and $g_5$, and they
should provide us important information for determining internal
nucleon spin structure. Next, nuclear structure functions are discussed.
From $F_3$ structure function measurements, valence-quark shadowing
should be clarified. Nuclear effects on the NuTeV
$sin^2 \theta_W$ anomaly are explained. We also comment on low-energy
neutrino scattering, which is relevant to current long-baseline
neutrino oscillation experiments.
\end{abstract}

\maketitle
\section{Introduction}\label{intro}

Nucleon structure has been investigated experimentally by various
scattering experiments. Now, the perturbative QCD is well understood.
The nonperturbative part is studied by theoretical models and lattice
calculations, and they are tested experimentally. Because of these efforts,
many aspects of the nucleon substructure are understood. However, there are
still missing points. For example, spin is a fundamental quantity,
and yet nucleon spin is poorly understood. We still do not know how the spin
is constituted in terms of quarks and gluons. Future neutrino facilities
should be able to provide important information on the internal
hadron structure including the spin.

High-energy neutrino reactions have been already used for investigating
the nucleon structure and determining fundamental constants such as
the running coupling constant $\alpha_s$ and weak-mixing angle
$sin^2 \theta_W$. From accurate neutrino deep inelastic scattering (DIS)
data, the structure functions, $F_1$, $F_2$, and $F_3$, have been extracted.
Future neutrino facilities, superbeams \cite{super} and neutrino factories
\cite{nufact}, will provide new insight into the hadron substructure.
Specialized talks are presented in the working  group 2 (WG2)
of this workshop, so that the details should be found in its summary
\cite{wg2sum} and presentations \cite{wg2-s, wg2-dis, wg2-sin, wg2-low}.
Neutrino beams are strong enough to allow proton and
polarized targets at the considered neutrino factories. Therefore,
the nucleon structure functions and the fundamental constants are
obtained without worrying about nuclear corrections. In addition,
it is important that polarized structure functions,
especially new functions $g_3$, $g_4$, and $g_5$, could be measured.
Using these polarized structure functions, we expect that
the internal nucleon spin structure will be precisely understood.

The future neutrino facilities are supposed to contribute also to
nuclear physics. In the present neutrino DIS, accurate measurements
have been done mainly for the nuclear target, iron, so that neutrino-nucleus
scattering data already exist. However, there is no accurate deuteron
or proton data for investigating nuclear corrections in neutrino
reactions by taking the ratio $\sigma_{\nu A}/\sigma_{\nu D}$.
Because the proton and deuteron cross sections should be accurately
measured at the future facilities, we could shed light on
the nuclear corrections. In particular, measurements of
the function $F_3$ will clarify the valence shadowing phenomenon.
On the other hand, we could investigate nuclear effects such as
Pauli exclusion and nucleon-nucleon correlation in the low-energy
scattering.

This paper consists of the following. The unpolarized and polarized
neutrino-nucleon scattering processes are explained, and then nuclear structure
functions are discussed. We also comment on low-energy neutrino scattering.
Finally, the discussions are summarized. 

\section{Unpolarized neutrino-nucleon scattering}\label{nu-n}

The cross section for unpolarized neutrino-nucleon DIS is calculated by
assuming a one-boson exchange process, and then the charged-current (CC)
cross section is expressed in terms of three structure functions,
$F_1$, $F_2$, and $F_3$:
\begin{equation}
\frac{d^2\sigma_{_{CC}}}{dx \, dy} = \frac{G_F^2 \, s}
                               {2 \, \pi \, (1+Q^2/M_W^2)^2} \,
     \bigg [ x \, y^2 \, F_1 + (1-y) \, F_2  
               \pm y \, (1-y/2) \, x \, F_3 \bigg ] .
\label{cross-cc}
\end{equation}
Here, $+$ and $-$ of $\pm$ indicate neutrino and antineutrino reactions,
respectively, $G_F$ is the Fermi coupling constant, $s$ is the center-of-mass
squared energy, $Q^2$ is defined by the momentum transfer $q$: $Q^2=-q^2$, 
and $M_W$ is the $W$ boson mass. The kinematical
variables $x$ and $y$ are defined by $x=Q^2/(2 M q^0)$ and $y=q^0/E$
with the nucleon mass $M$ and the initial neutrino energy $E$.
There are sum rules for these structure functions:
\begin{align}
   S_A & = \int_0^1 \frac{dx}{x} \, 
           \bigg [ \, F_2^{\bar \nu p} (x,Q^2) 
                    - F_2^{\nu p} (x,Q^2) \, \bigg ] 
         = 2 ,
\nonumber \\
   S_{Bj} & = \int_0^1 dx \, 
           \bigg [ \, F_1^{\nu n} (x,Q^2) 
                    - F_1^{\nu p} (x,Q^2) \, \bigg ] 
         = 1 - \frac{2}{3} \, \frac{\alpha_s (Q^2)}{\pi} 
             + \cdot\cdot\cdot
             + O(1/Q^2) ,
\\
   S_{GRS} & = \frac{1}{2} \int_0^1 dx \, 
           \bigg [ \, F_3^{\bar\nu p} (x,Q^2) 
                    + F_3^{\nu p} (x,Q^2) \, \bigg ] 
         = 3 \, \bigg [ 1 - \frac{\alpha_s (Q^2)}{\pi} 
             + \cdot\cdot\cdot \bigg ]
             + O(1/Q^2) .
\nonumber
\end{align}
These are called Adler, unpolarized Bjorken, and Gross-Llewellyn Smith
sum rules. There are perturbative QCD corrections to the last two sum rules,
and they have been investigated up to the $\alpha_s^4$ level \cite{nufact02-w3}.
Therefore, sum-rule measurements will provide valuable information
for an accurate determination of $\alpha_s$. Possible ambiguities come from
the higher-twist corrections $O(1/Q^2)$. Therefore, it is important to
understand twist-four corrections theoretically, and
such studies should be tested experimentally in the small-$Q^2$ region
at the future neutrino facilities. The details of these points are
summarized in the previous workshop \cite{nufact02-w3}.

The structure functions are expressed in terms of parton distribution
functions (PDFs). The CC cross section is calculated in the parton model
by using the current
\begin{equation}
J^\mu_{CC}  =  \bar u \, \gamma^\mu \, (1-\gamma_5) \,
       [ \, d \, \, cos \theta_c + s \, \, sin \theta_c \, ]
            +  \bar c \, \gamma^\mu \, (1-\gamma_5) \,
       [ \, s \, \, cos \theta_c - d \, \, sin \theta_c \, ]
\, ,
\label{eqn:jcc}
\end{equation}
where $\theta_c$ is the Cabbibo angle. Comparing the obtained cross section
with Eq. (\ref{cross-cc}), we have the leading-order (LO) expressions
for the structure functions in terms of the PDFs:
\begin{align}
2 \, x \, (F_1^{\nu p})_{CC}     = &
          (F_2^{\nu p})_{CC}     =   \, 2 \, x \, (\bar u + d + s + \bar c) ,
\nonumber \\
2 \, x \, (F_1^{\bar\nu p})_{CC} = &
          (F_2^{\bar\nu p})_{CC} =   \, 2 \, x \, (u + \bar d + \bar s + c) ,
\\
x \, (F_3^{\nu p})_{CC}      =  \, 2 \, x \, (-\bar u + d   & + s - \bar c) ,
\ \ \ \ 
x \, (F_3^{\bar\nu p})_{CC}  =  \, 2 \, x \, (u - \bar d   - \bar s + c) .
\nonumber 
\label{f123cclo}
\end{align}
\begin{wrapfigure}{r}{0.44\textwidth}
   \vspace{-0.3cm}
   \begin{center}
       \epsfig{file=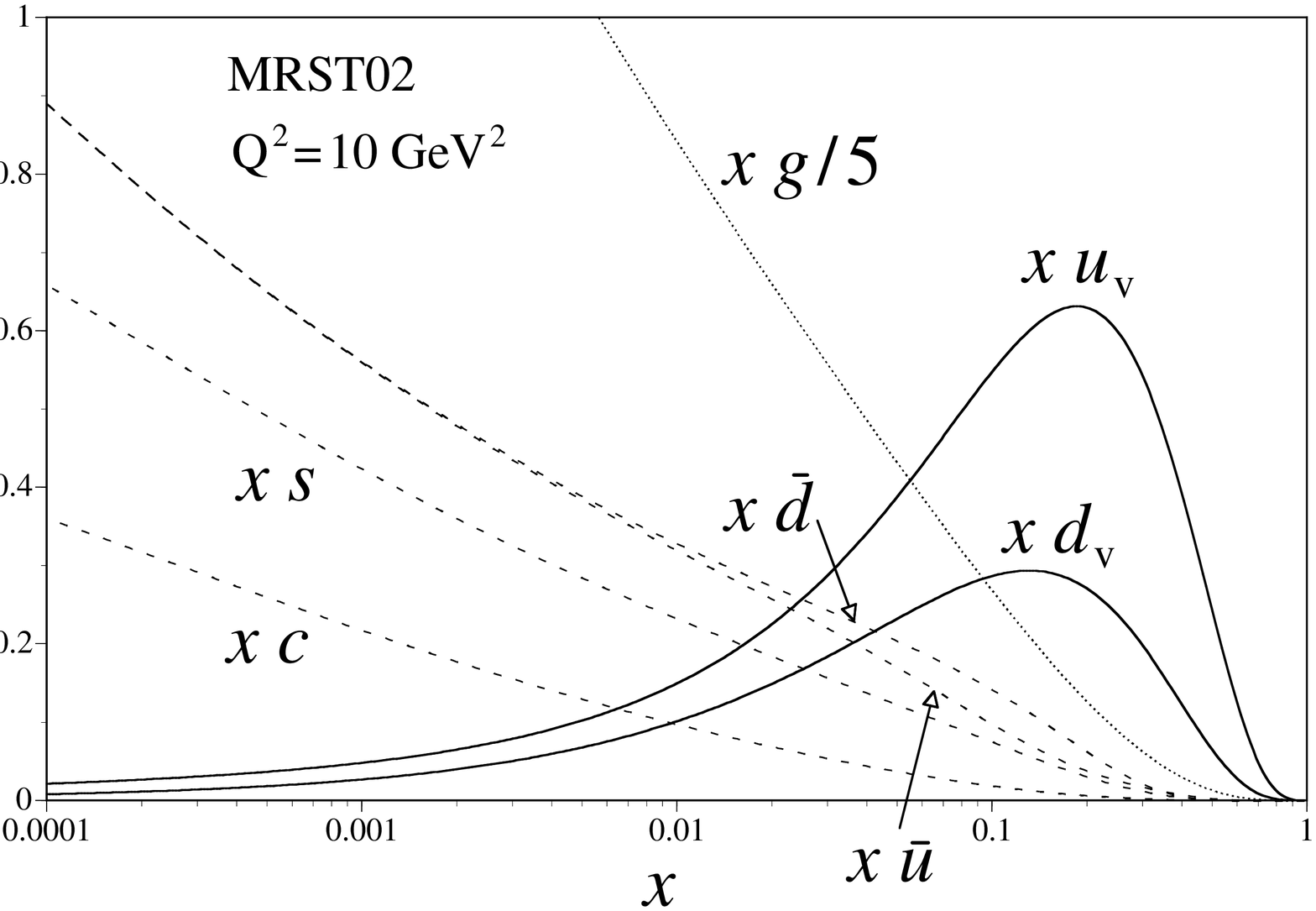,width=5.0cm} \\
       \vspace{-0.05cm}
       {\footnotesize {\bf FIGURE 1.}
            Parton distribution functions \\ at $Q^2$=10 GeV$^2$.}
   \end{center}
   \vspace{-0.3cm}
\end{wrapfigure}
\noindent
Neutron structure functions are obtained by using the isospin symmetry
for the PDFs. Parton-model expressions for neutral current (NC) structure
functions are not shown here, but they are found, for example, 
in Refs. \cite{roberts, neutrino}.

Using the neutrino DIS data together with other lepton and hadron scattering
data, we obtain the PDFs in the nucleon. The present situation is illustrated
in Fig. 1 \cite{pdfs}, where the MRST02 distributions are shown 
at $Q^2$=10 GeV$^2$ as an example. Because these distributions are rather
well determined, we had better focus on other aspects such as polarized
and nuclear PDFs at future neutrino facilities. 

\section{Polarized neutrino-nucleon scattering}\label{pol-nu-n}

\begin{wrapfigure}{r}{0.48\textwidth}
   \vspace{-0.3cm}
   \begin{center}
       \epsfig{file=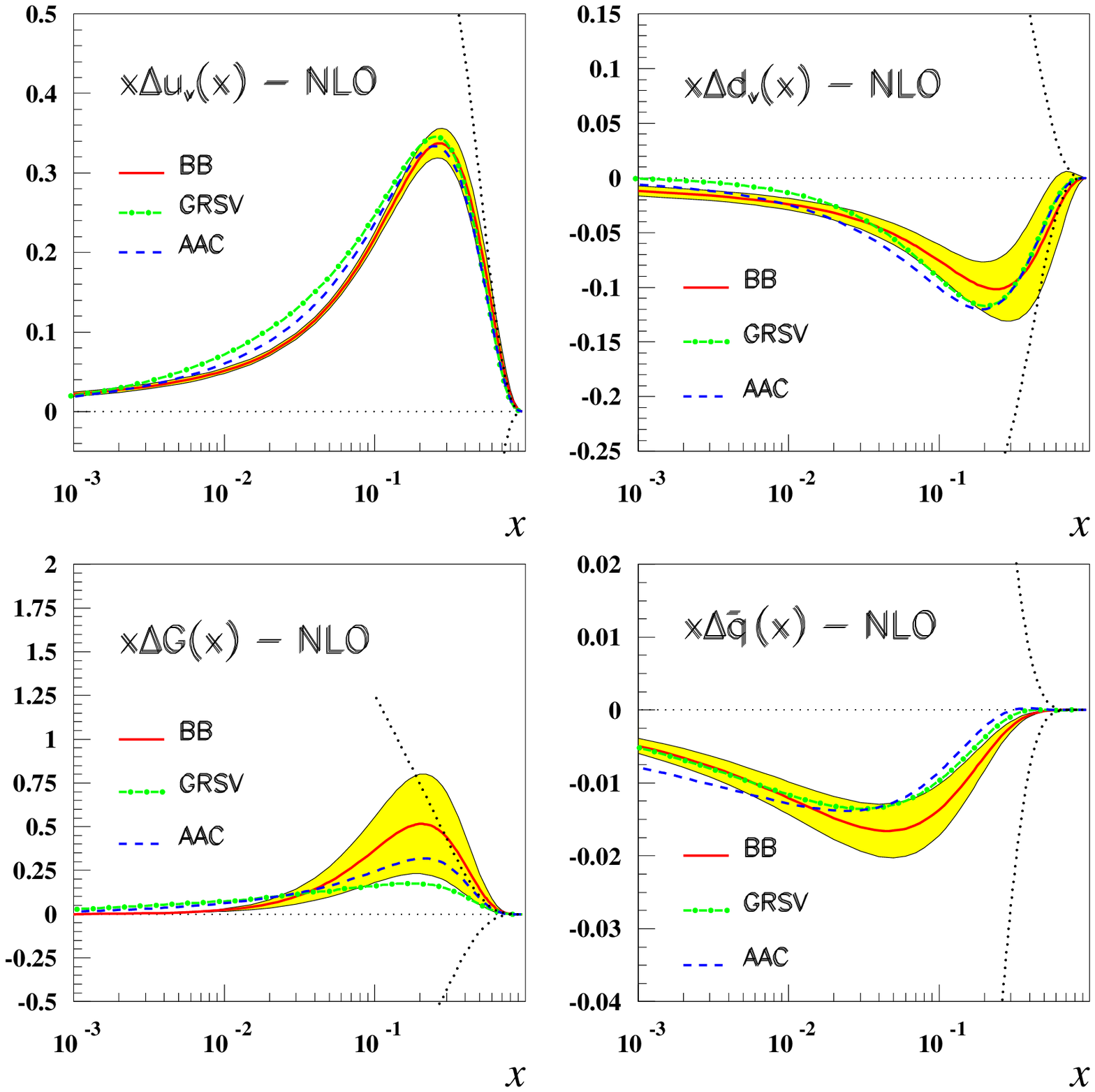,width=6.5cm} \\
       \vspace{0.1cm}
       {\footnotesize {\bf FIGURE 2.}
            Recent polarized PDFs \cite{polpdfs}.}
   \end{center}
   \vspace{-0.3cm}
\end{wrapfigure}

Polarized structure functions have been investigated by
electron and muon DIS. Current polarized PDFs are determined
by analyzing these data. Inclusive data are listed by the spin
asymmetry $A_1$, which is expressed 
$A_1 \cong  2 \, x \, (1+R) \, g_1 / F_2$, where $R$ is
the longitudinal-transverse structure function ratio 
and $g_1$ is a polarized structure function. The $g_1$ is given
by the polarized PDFs which are expressed by a number of parameters.
These parameters are determined by a $\chi^2$ analysis with the 
spin asymmetry data.

Recent analysis results are illustrated in Fig. 2 \cite{polpdfs},
where the polarized PDFs and their errors by Bl\"umlein and B\"ottcher
are shown as an example. The polarized valence-quark, antiquark,
and gluon distributions are shown. Three different parametrization
results are compared, and they agree each other except
for the gluon distribution. The error bands for
the valence-quark distributions are small; however, the error is large
especially for the gluon distribution. It indicates that the polarized
gluon distribution cannot be fixed at this stage.
 
The results may seem to indicate that the polarized PDFs are rather
well determined except for the gluon. However, there are important
points to be investigated. The overall magnitudes of the polarized
valence-quark distributions are fixed by low-energy seimileptonic decay
data with a flavor symmetric assumption for the antiquark distributions 
($\Delta \bar u=\Delta \bar d=\Delta \bar s$). Furthermore, the quark
spin content $\Delta \Sigma$ cannot be determined from the current
electron and muon DIS experiments although the analyses indicate a small
fraction $\Delta \Sigma=10-30 \%$. These issues could be clarified
by future neutrino DIS studies as explained in the following.

In addition to $g_1$ and $g_2$, there exist extra functions $g_3$,
$g_4$, and $g_5$ in neutrino reactions. There are various definitions
for $g_3$, $g_4$, and $g_5$ depending on researchers, so that one
should be careful in reading related papers. In the following, we
use the convention in Refs. \cite{nufact,g345,fmr}. The asymmetry
$\Delta\sigma$ is the difference between polarized cross sections:
$\Delta\sigma= \sigma_{\lambda_p=-1}-\sigma_{\lambda_p=+1}$,
where $\lambda_p$ is the proton helicity, and it is expressed as
\begin{equation}
\frac{d \Delta\sigma^{\lambda_\ell}}{dx \, dy}
= \frac{G^2_F}{\pi (1+Q^2/M_W^2)^2} \, \frac{Q^2}{x \, y} \,
   [ - \lambda_\ell \, x \, y \, (2-y) \, g_1
     - (1-y) \, g_4 - x \, y^2 \, g_5 ] \, ,
\end{equation}
for the CC process by neglecting $M^2/Q^2$ correction terms.
Here, $\lambda_\ell$ is the lepton helicity. In the parton model,
the leading-twist structure functions $g_1$, $g_4$, and $g_5$ are
expressed in terms of the polarized PDFs. The $g_4$ and $g_5$ are
related by the Callan-Gross type relation $g_4=2 x g_5$ in the LO,
and the CC structure functions $g_1$ and $g_5$ are expressed:
\begin{eqnarray}
& g_1^{\nu p} =  \Delta\bar u + \Delta d + \Delta s + \Delta\bar c,
\ \ \ \
& g_1^{\bar\nu p} = \Delta u +\Delta\bar d + \Delta\bar s + \Delta c,
\nonumber \\
& g_5^{\nu p} = \Delta\bar u - \Delta d -\Delta s +\Delta \bar c,
\ \ \ \
& g_5^{\bar\nu p} = -\Delta u + \Delta \bar d + \Delta \bar s - \Delta c .
\end{eqnarray}
It is important that the $g_1$ structure functions directly probe
the flavor singlet distribution:
$\Delta\Sigma (x)  = g_1^{(\nu+\bar\nu) p} 
 =\Delta u+\Delta\bar u +\Delta d+\Delta\bar d
+\Delta s+\Delta\bar s +\Delta c+\Delta\bar c $.
Therefore, the quark spin content issue could be clarified by
the neutrino scattering although the measured $x$ range is limited. 
In addition, combining the $g_5$ functions for the proton, we obtain 
$ g_5^{\nu p} + g_5^{\bar\nu p} = 
       - ( \Delta u_v + \Delta d_v ) 
       - ( \Delta s - \Delta \bar s ) - ( \Delta c - \Delta \bar c )$.
The $g_5$ functions are important for determining the polarized
valence-quark distributions.

\vspace{-0.2cm}
\noindent
\parbox[t]{0.46\textwidth}{
   \begin{center}
       \epsfig{file=deltau.ps,width=5.5cm} \\
   \end{center}
}\hfill
\parbox[t]{0.46\textwidth}{
   \begin{center}
       \epsfig{file=deltaub.ps,width=5.5cm} \\
   \end{center}
}
   \begin{center}
\vspace{-0.3cm}
       {\footnotesize {\bf FIGURE 3.}
          Feasibility studies for a neutrino factory \cite{fmr}. \\
          Expected $x(g_1^{\bar \nu p}-g_5^{\bar \nu p})/2$ and
          $x(g_1^{\nu p}+g_5^{\nu p})/2$ are shown.}
   \end{center}
\vspace{0.4cm}

Feasibility is studied for the European neutrino factory 
in Ref. \cite{fmr}, and some results are shown in Fig. 3.
Eight-year running of the neutrino factory with the butanol target is
assumed for estimating the errors. As shown in the figure, $\Delta u$ and 
$\Delta \bar u$ are the dominant contributions to the combinations
$g_1^{\bar\nu p}-g_5^{\bar\nu p}$ and $g_1^{\nu p}+g_5^{\nu p}$,
respectively, and they should be determined by the polarized reactions.
However, luminosity has be increased as much as possible for 
accurate measurements.

A recent HERMES analysis indicates a slightly positive $\Delta s (x)$
at small $x$ \cite{wg2-s} in contrast to the parametrization results
in Fig. 2. On the other hand, the polarized strangeness $\Delta s$ could be
investigated by other neutrino reactions \cite{wg2-s}. 
In elastic neutrino scattering, the axial vector form factor
$G_A (Q^2)$ can be measured. If non-strange contributions are known,
the strange part is extracted: $G_A^s (Q^2 \rightarrow 0)=\Delta s$.
At this stage, the analysis of BNL734 data indicates that $G_A^s$ is
consistent with zero. However, there is a proposal to measure it at Fermilab
by the FINeSE project \cite{wg2-s}. We expect that the strange spin
will be clarified by $G_A$ as well as the DIS experiments.

\section{Neutrino-nucleus scattering}\label{nu-nucl}

\begin{wrapfigure}{r}{0.44\textwidth}
   \vspace{-0.3cm}
   \begin{center}
       \epsfig{file=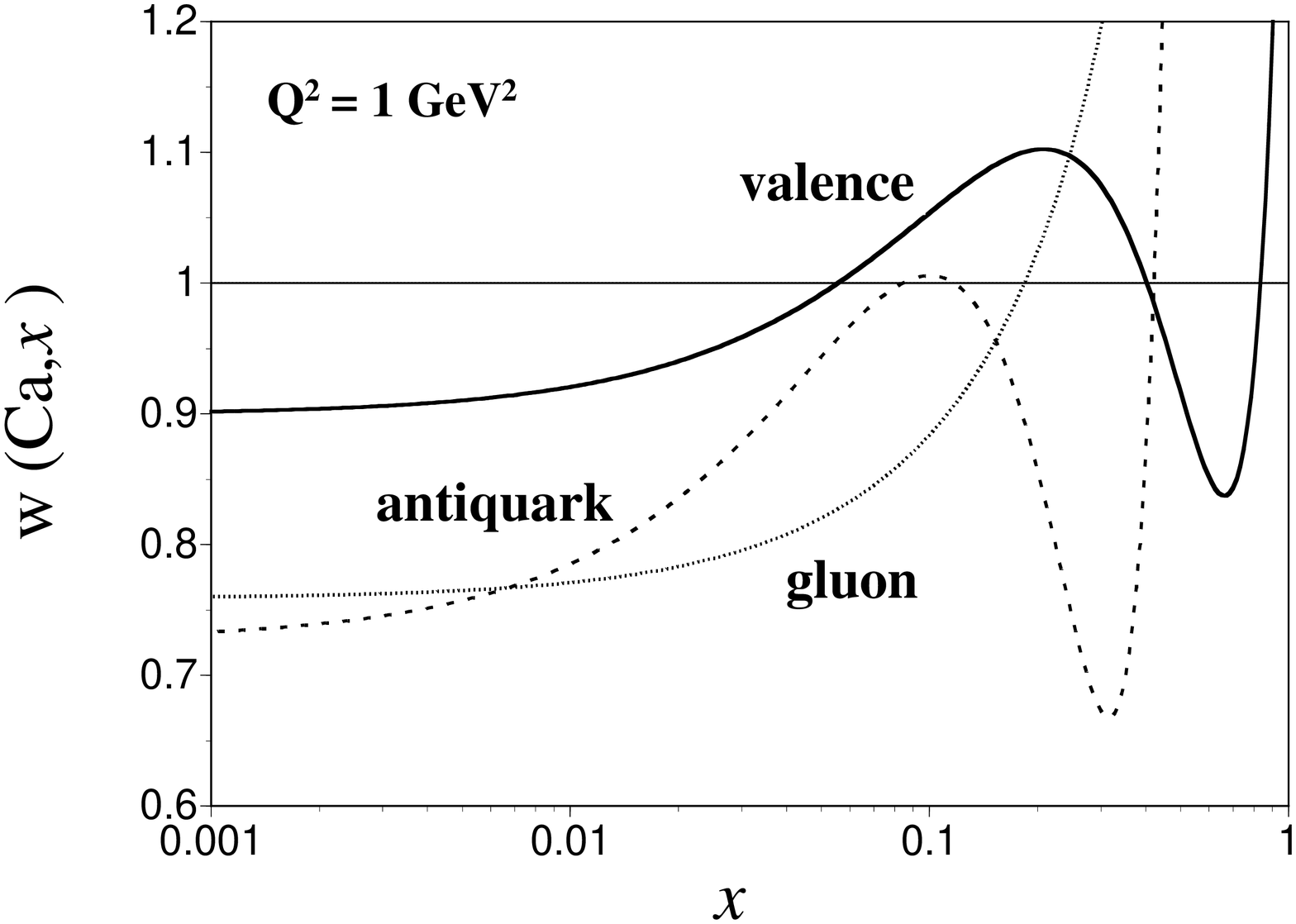,width=5.3cm} \\
       \vspace{-0.15cm}
       {\footnotesize {\bf FIGURE 4.}
            Nuclear modification of \\ the PDFs \cite{hkm}.}
   \end{center}
   \vspace{-0.3cm}
\end{wrapfigure}

Nuclear modification of the PDFs is investigated in lepton DIS and
high-energy hadron reactions. There are two major parametrizations,
EKRS \cite{ekrs} and HKM \cite{hkm}, for nuclear PDFs.
Current situation of the HKM studies is shown for the $^{40}Ca$
nucleus in Fig. 4, where $w(Ca,x)$ indicates nuclear modification.
A $\chi^2$ analysis has been made by using the data on
the structure-function ratios $F_2^{A}/F_2^{A'}$ and Drell-Yan
cross-section ratios $\sigma_{DY}^{pA}/\sigma_{DY}^{pA'}$.
The nuclear PDFs are expressed by a number of parameters,
which are then determined by the $\chi^2$ analysis with the data.
The valence-quark distributions are determined well in the medium-$x$ region.
Because the nuclear modification is negative in this region as shown
in Fig. 4, it is cancelled by the positive one at $x \approx 0.2$
so as to satisfy the charge and baryon-number conservations.
However, these conservations do not pose a strong constraint
at small $x$, so that the small-$x$ modification is not
obvious for the valence-quark distributions, and it should be tested by
future neutrino DIS measurements. The antiquark distributions are fixed 
by the observed $F_2$ shadowing at small $x$ and the Drell-Yan
data at $x\sim 0.1$; however, the medium-$x$ behavior is not obvious
unless new data are obtained. It is difficult to determine the nuclear
gluon distributions at this stage.

Because the iron target has been used in neutrino scattering, there are
already neutrino-nucleus scattering data. However, it is not possible to
investigate nuclear modification due to the lack of accurate deuteron data.
At future neutrino facilities, proton and deuteron measurements will become
possible, so that the nuclear modification could be investigated.
In particular, the $F_3$ structure function is specific
in the neutrino reaction. Although the $F_2$ shadowing is investigated
well in the electron and muon scattering, $F_3$ shadowing has not been
studied at all. The function $F_3$ provides information on the valence-quark
distribution: $(F_3^{\nu N}+F_3^{\bar\nu N})_{CC}/2 \cong u_v+d_v$.
The NuMI project \cite{numi} and ultimately the neutrino factories will
provide data for the difference between the $F_2$ and $F_3$ shadowing
modifications, so that this issue will become clear in future.


We comment on a possible nuclear modification of the longitudinal-transverse
structure function ratio $R$. It is sometimes called ``HERMES effect".
The effect was suggested by the HERMES collaboration in 2000 \cite{hermes};
however, it was not observed in a CCFR/NuTeV analysis of neutrino data
\cite{ccfr01} and also in a subsequent HERMES re-analysis with careful
radiative corrections \cite{hermes}.
Nonetheless, there could be a nuclear modification at large $x$ with
small $Q^2$, which is not the observed region by these experiments.
A physics origin is the admixture of longitudinal and transverse
nucleon structure functions in a nucleus due to nucleon Fermi motion
\cite{ek03}. Such an effect could be investigated by JLab experiments
\cite{jlab-r} and possibly by future neutrino reactions.

\subsection{$\mathbf{sin^2 \theta_W}$ anomaly from a nuclear physicist's
            point of view}\label{sint}

The NuTeV collaboration reported anomalously large weak mixing angle:
$sin^2 \theta_W = 0.2277 \pm 0.0013 \, \text{(stat)} 
                         \pm 0.0009 \, \text{(syst)}$ \cite{wg2-sin,nutev02}
in comparison with a global analysis result
$sin^2 \theta_W = 0.2227 \pm 0.0004$ without neutrino-nucleus
scattering data \cite{lep01}. In the WG2 of this workshop, there are
discussions on the NuTeV result and future experimental
studies by parity-violating DIS and M\o ller scattering \cite{wg2-sin}.
Although there may be new physics \cite{new} behind this difference,
it is more natural to seek a mechanism in nuclear corrections
of the target iron.

The Paschos-Wolfenstein relation, 
$ R^-  = ( \sigma_{NC}^{\nu N}  - \sigma_{NC}^{\bar\nu N} ) /
         ( \sigma_{CC}^{\nu N}  - \sigma_{CC}^{\bar\nu N} )
        =  \frac{1}{2} - sin^2 \theta_W $,
plays an important role for extracting $sin^2 \theta_W$ from 
neutrino and antineutrino scattering data. This relation is valid for
the isoscalar nucleon, but there are four correction factors in a nucleus:
$R_A^-   =  \frac{1}{2} - sin^2 \theta_W  
    +O(\varepsilon_v)+O(\varepsilon_n)+O(\varepsilon_s)+O(\varepsilon_c)$.
Here, $O(\varepsilon)$ indicates a correction of the order of $\varepsilon$,
and the detailed expressions should be found in 
Refs. \cite{sk02,kulagin,nucl-sinth}. The correction factors are defined by
\begin{alignat}{2}
  \varepsilon_n (x) & = \frac{N-Z}{A} \,
                      \frac{u_v(x)-d_v(x)}{u_v(x)+d_v(x)} \, , \ \ \ \ \ \ \ 
& \varepsilon_v (x) &  = \frac{w_{d_v}(x,A)-w_{u_v}(x,A)}
                             {w_{d_v}(x,A)+w_{u_v}(x,A)} \, , 
\nonumber \\
  \varepsilon_s (x) & = \frac{s^A(x)-\bar s^A(x)}
                         {w_v(x,A) \, [u_v(x)+d_v(x)]} \, , \ \ \ \ \ \ \ 
& \varepsilon_c (x) & = \frac{c^A(x)-\bar c^A(x)}
                         {w_v(x,A) \, [u_v(x)+d_v(x)]} \, , 
\end{alignat}
where $w_{u_v}$ and $w_{d_v}$ indicate the nuclear modifications of
up- and down-valence quark distributions, and $w_v$ is defined by
$w_v=(w_{u_v}+w_{d_v})/2$.
The function 
$\varepsilon_n$ comes from the non-isoscalar nature of the nucleus, 
$\varepsilon_v$ is related to the nuclear modification difference 
                between $u_v$ and $d_v$, and
$\varepsilon_s$ ($\varepsilon_c$) is proportional to the 
difference between $s$ and $\bar s$ ($c$ and $\bar c$).
The charm correction is expected to be small. The strange correction
is also found to be a small effect according to a NuTeV estimate,
and it tends to increase the deviation \cite{nucl-sinth}.
The valence-quark correction $\varepsilon_n$ was found to be small
according to model estimates \cite{sk02} although it should be
tested by future experiments.
The isovector correction $\varepsilon_n$ was included in the NuTeV
analysis. It was later investigated in Ref. \cite{kulagin}; however,
NuTeV kinematical effects may reduce such a contribution. Therefore,
it seems that the nuclear effects \cite{sk02,kulagin,nucl-sinth} are
not enough for explaining the whole deviation at this stage.

\section{Comments on low-energy neutrino scattering}\label{low-nu}

We have discussed high-energy neutrino reactions; however, current
long baseline neutrino experiments have been done in the low-energy region.
In order to understand the neutrino oscillation parameters in a few
percent accuracy, the neutrino cross section should be understood
accurately as well \cite{wg2-low}. There are two important factors.
One is to understand the neutrino interaction with the $^{16}O$ nucleus,
another is to describe the cross sections in both DIS and resonance regions.
There is a dedicated workshop for this topic, so that the details
are found in its web page \cite{nuint012}.

\begin{wrapfigure}{r}{0.46\textwidth}
   \vspace{-0.6cm}
   \begin{center}
       \epsfig{file=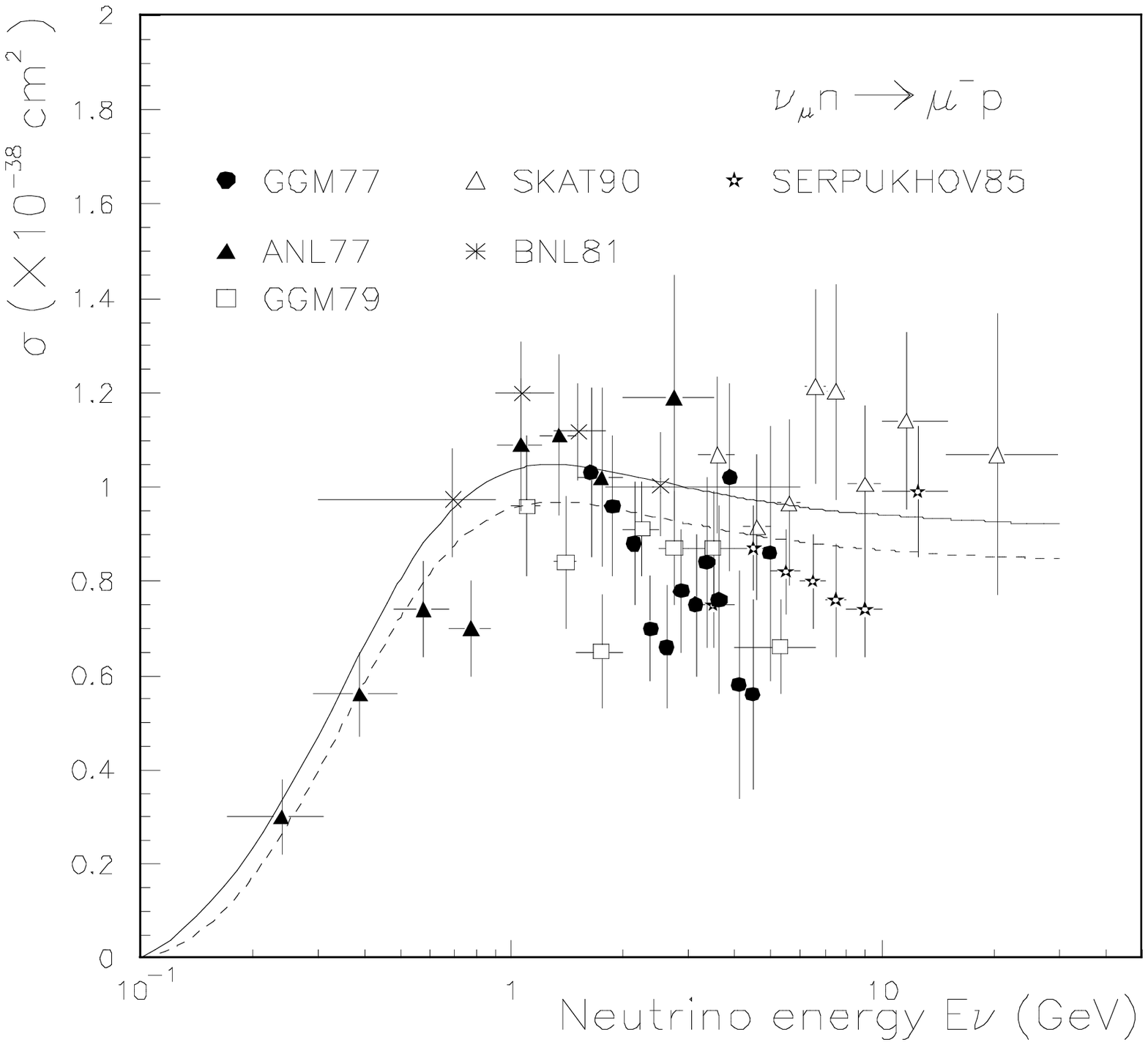,width=6.5cm} \\
       \vspace{-0.30cm}
       {\footnotesize {\bf FIGURE 5.}
            Quasi-elastic cross section \cite{sakuda}.
            The dashed curve includes exclusion effects.}
   \end{center}
   \vspace{-0.3cm}
\end{wrapfigure}
First, nuclear corrections should be accurately taken into account
\cite{sakuda}. At high energies, they are expressed in terms of
the nuclear PDF modifications. At low energies, the corrections
include the effects of nuclear binding, Fermi motion, Pauli exclusion,
and nucleon-nucleon ($NN$) correlation. For example, a final-state nucleon
in a neutrino reaction suffers from the exclusion effect due to
the existence of other nucleons. Such effects modify the small $Q^2$ part
of the cross section significantly. If the cross section is shown
as a function of neutrino energy, the exclusion effect is typically 8\%
as shown in Fig. 5  \cite{sakuda}. From the figure, it is also obvious
that the cross section is not accurately measured, and this fact makes
it difficult to determine the oscillation parameters accurately. 
Furthermore, the $NN$ correlation mechanism gives rise to a large momentum
tail beyond the Fermi momentum, and it also modifies the cross section
significantly. All of these nuclear corrections should be understood
clearly for the precise neutrino-oscillation physics.

Second, an appropriate model should be studied for describing
the cross section smoothly from the DIS to the resonance region
because the neutrino data could contain both contributions.
It is shown in Ref. \cite{by} that a simple change of the scaling
variable [$x_w=x (Q^2+0.624)/(Q^2+1.735 x)$] could describe the 
measured $F_2$ structure functions fairly well even in the small
$Q^2$ region ($Q^2$=0.07, 0.22, and 0.85 GeV$^2$). 
Such a simple prescription could be also applied to the neutrino cross
sections for the description in both low- and high-energy regions.

\vspace{-0.1cm}
\section{Summary}\label{summary}

Future superbeam and neutrino factories provide us a unique opportunity
for investigating nucleon substructure, which cannot be studied
by other lepton and hadron probes. Nucleon spin structure will be
clarified by the leading-twist structure function $g_1$ and $g_5$.
The valence-quark shadowing will be investigated by the $F_3$ structure
functions for nuclei. We pointed out that these studies together
with low-energy nuclear structure studies affect the long-baseline
experiments as nuclear corrections. The hadron-structure studies
are important, for example, for finding a quark-gluon plasma signature
and any exotic signature beyond the current physics framework.
The future neutrino facilities should play an important role
in establishing the hadron-structure physics.

\vspace{-0.1cm}
\begin{theacknowledgments}
S.K. would like to thank Y. Kuno for motivating him to hadron-structure
studies at the neutrino factories. He thanks the Elsevier Science for
permitting him to quote Figs. 2, 3, and 5 directly from its publications
\cite{polpdfs,fmr,sakuda}. He was supported by the Grant-in-Aid for
Scientific Research from the Japanese Ministry of Education, Culture,
Sports, Science, and Technology. 
\end{theacknowledgments}

\vspace{-0.2cm}
\bibliographystyle{aipprocl} 


\end{document}